\documentclass{article}
    \usepackage{spconf}
    \title{On Optimizing Image Codecs for VMAF NEG: \\
            Analysis, Issues, and a Robust Loss Proposal}

        \name{Florian Fingscheidt$^{*}$$^{\dagger}$, Alexander Karabutov$^{*}$, Panqi Jia$^{*}$, Elena Alshina$^{*}$, Jörn Ostermann$^{\dagger}$}
            \address{$^{*}$ Huawei Technologies Düsseldorf GmbH,\hspace{1cm} $^{\dagger}$ Leibniz University Hannover \\
            \hspace{2.5cm} Riesstraße 25, \hspace{3.2cm} Institute for Information Processing \\
            \hspace{1cm} 80992 Munich, Germany \hspace{3cm} 30167 Hannover, Germany}

\usepackage{amsmath,graphicx}
\usepackage{amssymb}
\usepackage[dvipsnames]{xcolor}
\usepackage{multirow} % For vertically merging cells
\usepackage{caption}
\usepackage{subcaption} 
\usepackage{lipsum}
\usepackage[hyphens]{url}
\usepackage{hyperref}
\hypersetup{breaklinks=true}

\makeatletter
\renewcommand\p@subfigure{\thefigure}

\makeatother

\begin{document}

\ninept%            <-- OPTIONAL, for nine pt only
\maketitle
\begin{abstract}
    The VMAF (video multi-method assessment fusion) metric for image and video coding recently gained more and more popularity as it is supposed to have a high correlation with human perception. This makes training and particularly fine-tuning of machine-learned codecs on this metric interesting. However, VMAF is shown to be attackable in a way that, e.g., unsharpening an image can lead to a gain in VMAF quality while decreasing the quality in human perception.
    A particular version of VMAF called VMAF NEG has been designed to be more robust against such attacks and therefore it should be more useful for fine-tuning of codecs. In this paper, our contributions are threefold. First, we identify and analyze the still existing vulnerability of VMAF NEG towards attacks, particulary towards the attack that consists in employing VMAF NEG for image codec fine-tuning. Second, to benefit from VMAF NEG's high correlation with human perception, we propose a robust loss including VMAF NEG for fine-tuning either the encoder or the decoder. Third, we support our quantitative objective results by providing perceptive impressions of some image examples.
\end{abstract}

\section{Introduction}
\label{sec:intro}
In recent years, neural network-based image codecs gained popularity within the coding community and show superior performance compared to classical codecs such as JPEG \cite{JPEG}, JPEG2000 \cite{JPEG2000}, HEVC \cite{HEVC}, and VVC \cite{VVC}. Such neural network-based codecs are, e.g., Cheng et al.\ (\texttt{Cheng2020}) \cite{Cheng2020}, Minnen et al.\ (\texttt{Mbt2018}) \cite{Mbt2018}, Liu et al.\ (\texttt{TCM}) \cite{TCM}, and \texttt{JPEG AI} \cite{JPEG-AI}, the latter being the first neural network-based image coding standard. 

VMAF \cite{VMAF} is an image and video quality metric with high correlation to human visual perception \cite{proof_vmaf_1}. Jenadeleh et al. \cite{proof_vmaf_2} showed that VMAF NEG \cite{VMAF_NEG}, which is supposed to be more robust against tuning and attacks \cite{VMAF_NEG}, got the highest correlation with human perception on JPEG AI \cite{JPEG-AI}. Accordingly, it could be advantageous to fine-tune an image codec with VMAF NEG to enhance the codec's visual perception, while maintaining about the same rate as before.

In this work, we reveal pitfalls when fine-tuning on VMAF NEG either the encoder (for improved services from content providers) or the decoder (for consumer device software updates) of an image codec. We present how to avoid these pitfalls by combining the VMAF NEG loss with other distortion losses to both stabilize the training and to obtain better subjective results. We backup our findings by presenting quantitative objective and qualitative subjective experimental results on three different neural image codecs.

% In \cite{proof_vmaf_1} they proofed that VMAF got a high correlation with human perception.

% In \cite{proof_vmaf_2} they proofed that VMAF NEG got the best correlation with visual perception on the JPEG AI codec (which is the first ever AI Image coding standard).

\section{Related Works}
\label{sec:related_works}

Zvezdakova et al.\ \cite{hacking_vmaf_1} showed that VMAF \cite{VMAF} is attackable by pre-processing the images with  unsharp masks and histogram equalization to enhance the VMAF score. Siniukov at al.\ \cite{hacking_vmaf_2} showed that not only VMAF, but also its tuning-resistant version VMAF NEG \cite{VMAF_NEG} is attackable by pre-processing videos with different strategies. Both, the  VMAF and VMAF NEG scores rise whith some of the pre-processing approaches, while the visual quality drops or remains the same, indica\-ting that both VMAF and VMAF NEG are attacked by these approches.

Aistov et al.\ \cite{VMAF-TORCH} presented differentiable versions of VMAF and VMAF NEG in the \texttt{VMAF-torch} software package. It is open-source and built to be used with \texttt{PyTorch}. Due to its differentiability, it is possible to calculate gradients through it and therefore it opens the opportunity to train or fine-tune image and video codecs on this version of VMAF---without the need of any proxy metric network.

Chen et al.\ \cite{vmaf_proxy} presented a proxy and learning framework for training neural image codecs on perceptual losses including VMAF. The codec is learned in an adversarial way, that performs alternating updates of both the codec and the proxy network to prevent codec and proxy network degradation. \textit{Different from that, we don't need to adversarially train a proxy}, since VMAF NEG from \texttt{VMAF-torch} \cite{VMAF-TORCH} is already differentiable, \textit{thus eliminating the need for a proxy network and giving us the benefit to train directly on the VMAF loss} provided by \texttt{VMAF-torch}. Also different from Chen et al.\ \cite{vmaf_proxy}, we don't train the entire codec, but we merely fine-tune either the encoder or the decoder in separate experiments for different deployment use cases.

% Inspired by Chen's work \cite{vmaf_proxy}, Lavrushkin et al.\ \cite{vmaf_adv_attack} created a training pipeline to be more robust against adversarial attacks by generating distorted images from adversarial attacks and creating a whole new dataset with adversarially distorted images and other distorted images that fooled the VMAF proxy and are able to fool the original VMAF. With this partly adversarially generated dataset and their proposed training pipeline, they were able to train a more robust proxy and therefore a better performing codec w.r.t.\ visual perception. The drawback of this approach is that this type of defense method against adversarial attacks has been stated to be only suitable for optimizing towards non-differentiable metrics. Different from us they used a preprocessing U-Net\cite{U_Net} and the not learning based x264 \cite{x264} instead of a learning based codec like we do. They only investigated the preprocessing where we investigate at encoder and decoder fine-tuning.
% In our work, we aim at the same goal of being more resistant in training, but we overcome the need of an adversarial training as our version of VMAF NEG is differentiable and we prevent the above mentioned pitfalls by using additional guidance from other quality metrics without the need of designing a new dataset.

Inspired by Chen’s work \cite{vmaf_proxy}, Lavrushkin et al.\ \cite{vmaf_adv_attack} created a training pipeline to make a VMAF proxy more robust against adversarial attacks. They generated distorted images using adversarial methods and built a new dataset containing adversarially distorted samples and other distortions that fooled both the proxy and the original VMAF. Using this adversarially enriched dataset, they trained a more robust non-differentiable VMAF proxy---but not a codec. The drawback of this approach is that \textit{this type of defense method against adversarial attacks has been stated to be only suitable for improving a non-differentiable metric} \cite{vmaf_adv_attack}. Different from us, they used a preprocessing U-Net \cite{U_Net} together with the non-learning-based x264 \cite{x264}, \textit{whereas we investigate fine-tuning of either encoder or decoder in a fully learned codec}.

% von Siniukov at al. [17] abgrenzen
In our work, we aim at the same goal of being more robust to attacks, but unlike \cite{vmaf_adv_attack}, we address robustness during fine-tuning and not during evaluation. More important, we overcome the need of an adversarial training as our version of VMAF NEG is differentiable and we prevent the above mentioned pitfalls by using additional guidance from other quality metrics without the need of designing a new dataset. Unlike Siniukov at al.\ \cite{hacking_vmaf_2}, our attacks are learned and not designed. Moreover, we don't try to attack VMAF NEG as much as possible, but we rather strive towards limiting the attackability of VMAF NEG in our loss function.

% In \cite{hacking_vmaf_1} they hack VMAF with two types of preprocessing.

% In \cite{hacking_vmaf_2} they hack VMAF and VMAF NEG with many different preprocessings.

% In \cite{vmaf_proxy} they train a proxy for VMAF to get a differentiable loss from it.

% Main concurrence paper: \cite{vmaf_adv_attack} for distancing see green marks: they train a proxy for GAN like training and they only claim their solution for non differentiable metrics like VMAF and not for differentiable ones like VMAF Torch.(They extend on the work of \cite{vmaf_proxy})

% In \cite{VMAF-TORCH} they show a differentiable version of VMAF and VMAF NEG called VMAF-torch, We use that.

\section{Method}
\label{sec:method}

To investigate the fine-tuning of our selected codecs, we divide the experiments into two applications: the first application builds upon \textit{fine-tuning only the encoder} along with a potential hyper-encoder, but we fix the decoder during training. This would only change the transmitter side, so that, e.g., image clouds and image databases could use the fine-tuned encoder while interoperability with existing receivers is preserved. The second application builds upon \textit{fine-tuning only the decoder}, while fixing the encoder, so that images don't have to be coded again, and each receiving device could use the new decoder as a software update.

To overcome VMAF NEG governing the codec during fine-tuning into directions of more distortion by being attacked, we propose a mixed loss approach where MSE and/or MS-SSIM guide the fine-tuning. For MS-SSIM, we adopt the implementation from the \texttt{pytorch\_msssim} toolbox \cite{ms_ssim}. This avoids getting fooled by VMAF NEG. Our proposed loss function is:

\begin{equation}
\begin{split}
    \mathcal{L} = R + \lambda \cdot \big(\alpha \cdot \text{MSE} + \beta \cdot (1 - \text{MS-SSIM})& \cdot \beta' \\ + \gamma \cdot (100 - \text{VMAF NEG})& \cdot \gamma'\big),
\end{split}
\label{ex:loss_function}
\end{equation}

where VMAF NEG is the mean of VMAF NEG on Y, U, and V channel and MSE is calculated on the RGB space as it it the native space of our codecs. Here, $R$ is the rate that we obtained from \texttt{CompressAI}'s bpp loss calculation, which is the total negative log-likelihood (in bits) of all latent symbols, divided by the number of pixels. We also use $\beta'$ and $\gamma'$ to scale our reversed distortion metrics $(1 - \text{MS-SSIM})\in[1,0]$ and $(100 - \text{VMAF NEG})\in\mathbb{R}$ \textit{individually for each codec} to match the MSE value range on validation data before fine-tuning and then keep those scaling factors $\beta',\gamma' > 0$ fixed throughout the fine-tuning. Afer having specified $\beta'$ and $\gamma'$, we determine by $\alpha$, $\beta$, $\gamma$ $\in[0,1]$ how much weight we assign to each distortion metric. Note that we always choose $\alpha + \beta + \gamma = 1$, allowing to easily interprete the percentage of each metric to the total distortion loss (\ref{ex:loss_function}).

\section{Experimental setup}
\label{sec:setup}

\subsection{Codecs}
\label{sec:codecs}
We investigate three image codecs, namely the \texttt{CompressAI} \cite{compressai} implementations of Cheng at al.\ \cite{Cheng2020} called Cheng2020-anchor (here: \texttt{Cheng2020}) and \texttt{Mbt2018} \cite{Mbt2018}, along with \texttt{TCM} \cite{TCM} which is a mixed convolution and attention network. All codecs realize a joint autoregressive and hierarchical prior architecture \cite{Mbt2018}.

\subsection{Datasets}
\label{sec:datasets}
The baseline training and all fine-tunings are conducted on the Vimeo90k dataset \cite{Vimeo90k}, using only the first image of each sequence contained therein. We use the official Vimeo90k training split for training and the test split for evaluation during training. For testing and demonstrating the effects of hyper-parameters, we use the Kodak dataset \cite{Kodak}.

\subsection{Training and Fine-Tuning Setup}
\label{sec:fine_tuning_setup}
We keep the learning rate fixed throughout all of our experiments at a value of $0.0001$. We learn on randomly cropped image patches of size $256\times256$. As our baseline, we use the pre-trained models at their quality level 3 and append a sigmoid activation function to clip the decoded images to the range $[0,1]$. We first train these codecs for 10 epochs with $\alpha=1$, $\beta=\gamma=0$, i.e., optimizing for MSE. We choose $\lambda=0.01$ for \texttt{Cheng2020} and \texttt{Mbt2018}, and $\lambda=0.025$ for \texttt{TCM}, thereby ensuring that the codecs are well adapted to the sigmoid activation function. We call these networks our baselines. 
Each of our employed fine-tunings, if not stated otherwise, consists of 10 further epochs employing Vimeo90k training and validation splits. 

The codec-individual scaling factors in our loss function (\ref{ex:loss_function}) are as follows: \texttt{Cheng2020} [$\beta'$, $\gamma'$] = [1600, 1.6], \texttt{Mbt2018} [1500, 2.0], \texttt{TCM} [1300, 1.5]. These scaling factors where obtained on the Vimeo90k validation dataset to match value ranges of MSE, MS-SSIM, and VMAF NEG.

\subsection{Bitrate Matching}
\label{sec:bitrate_matching}
Whenever we fine-tuned the \textit{encoder} and therefore possibly changed the bitrate of our codec, we performed bitrate matching to obtain comparable results for that particular encoder architecture throughout our various experiments. The bit\-rate matching is conducted by tuning the trade-off hyper-parameter $\lambda$ between distortion and bitrate $R$ to match the bitrate of the respective baseline codec. We tested the matching of bitrates on the Kodak dataset to assure that the images of Kodak are coded with similar bitrates. Throughout all of our experiments, we only allow a maximum rate deviation of the average rate by $\pm 1\%$ on Kodak, between a baseline ($\alpha=1, \beta=\gamma=0$) and proposed codec setting based on (\ref{ex:loss_function}) using $\beta \not=0$ and/or $\gamma\not=0$.

\section{Experimental Results}
\label{sec:experimental_results}

\begin{table}[t]
    \centering
    \begin{tabular}{@{}c@{\hspace{4pt}}l|r|r|r|r}
    & Fine-tuning & \multicolumn{1}{c|}{PSNR} & \multicolumn{3}{c}{VMAF NEG} \\
    \multirow{3}{*}{\rotatebox{90}{\texttt{TCM} \hspace{0.8cm} \texttt{Mbt2018} \hspace{0.15cm} \texttt{Cheng2020} \hspace{0.35cm}}} && \multicolumn{1}{c|}{(dB)} & \multicolumn{1}{c|}{Y} & \multicolumn{1}{c|}{U} & \multicolumn{1}{c}{V} \\
         \hline
         %Cheng2020 &&\\
         & None (= baseline) & 31.55 & 79.78 & 73.82 & 66.06 \\
         & ENC w/ MSE & 31.59 & 79.70 & 74.92 & 65.84 \\
         & DEC w/ MSE & 31.60 & 79.61 & 74.59 & 66.04 \\
         & ENC w/ VMAF NEG & \textcolor{red}{29.22} & \textcolor{green}{82.83} & \textcolor{green}{82.84} & \textcolor{green}{78.02} \\
         & DEC w/ VMAF NEG & \textcolor{red}{13.86} & \textcolor{green}{85.10} & \textcolor{green}{79.37} & \textcolor{green}{72.69} \\
         \hline
         %Mbt2018 &&\\
         & None (= baseline) & 31.68 & 78.75 & 71.99 & 64.81 \\
         & ENC w/ MSE & 31.68 & 78.76 & 73.10 & 64.46 \\
         & DEC w/ MSE & 31.71 & 78.72 & 73.03 & 65.24 \\
         & ENC w/ VMAF NEG & \textcolor{red}{29.94} & \textcolor{green}{82.48} & \textcolor{green}{81.68} & \textcolor{green}{77.08} \\
         & DEC w/ VMAF NEG & \textcolor{red}{20.91} & \textcolor{green}{82.03} & \textcolor{green}{79.14} & \textcolor{green}{72.49} \\  
         \hline
         %TCM &&\\
         & None (= baseline) & 35.20 & 87.23 & 82.23 & 76.94 \\
         & ENC w/ MSE & 35.21 & 87.01 & 81.33 & 76.90 \\
         & DEC w/ MSE & 35.28 & 87.42 & 82.42 & 76.99 \\
         & ENC w/ VMAF NEG & \textcolor{red}{31.91} & \textcolor{green}{89.37} & \textcolor{green}{87.51} & \textcolor{green}{84.59} \\
         & DEC w/ VMAF NEG & \textcolor{red}{13.61} & \textcolor{green}{89.88} & \textcolor{green}{86.48} & \textcolor{green}{83.14} \\  
    \end{tabular}
    \caption{Performance after \textbf{encoder or decoder fine-tuning}, with the distortion loss being either MSE or VMAF NEG. Hyperparameter $\lambda$ is tuned to get matching rates within each codec architecture. All metrics are calculated on the Kodak dataset. PSNR in (dB) on RGB, VMAF NEG on Y, U, V channels. Poor results in \textcolor{red}{red}, strong results in \textcolor{green}{green}.}
    \label{tab:first_fine_tuning_decoder}
\end{table}

\begin{figure}[t]
    \centering
    \begin{subfigure}[b]{0.235\textwidth}
        \includegraphics[width=\linewidth]{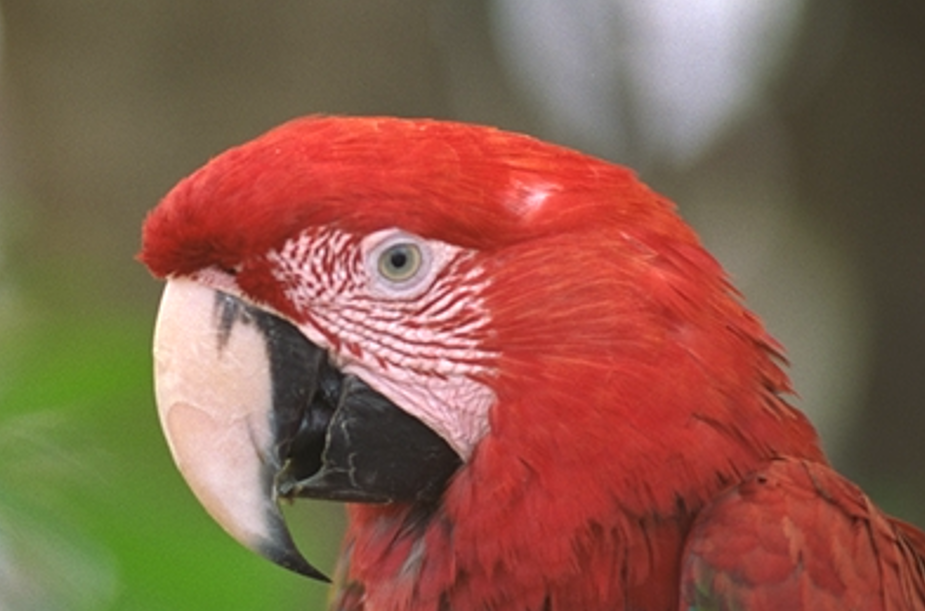}
        \caption{\centering Original\\\phantom{PSNR}}
        \label{fig:VMAF_YUV_fine_tuning_img_original}
    \end{subfigure}
    \begin{subfigure}[b]{0.235\textwidth}
        \includegraphics[width=\linewidth]{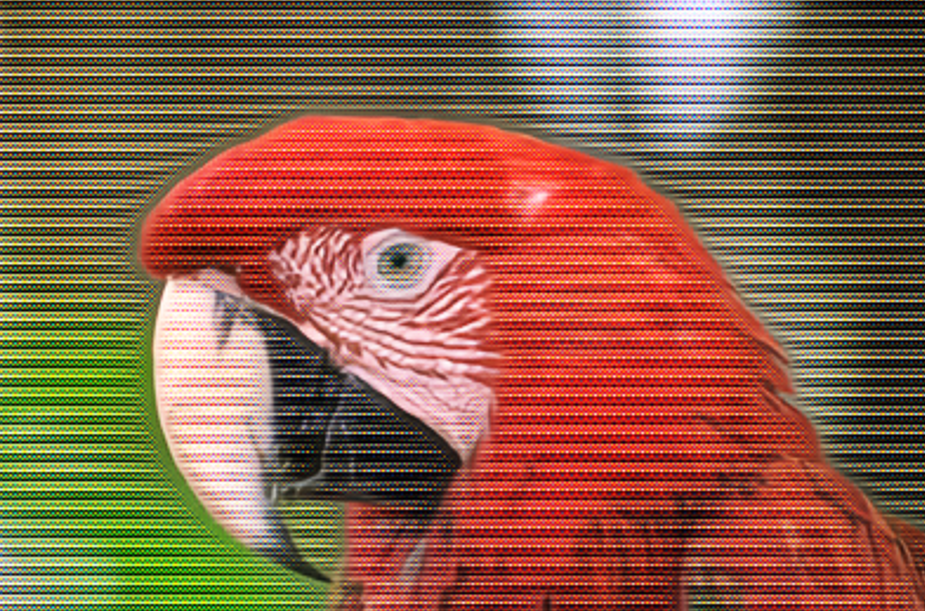}
        \caption{\centering \texttt{Cheng2020} \\ PSNR = 12.76 dB}
        \label{fig:VMAF_YUV_fine_tuning_img_cheng2020}
    \end{subfigure}
    \begin{subfigure}[b]{0.235\textwidth}
        \includegraphics[width=\linewidth]{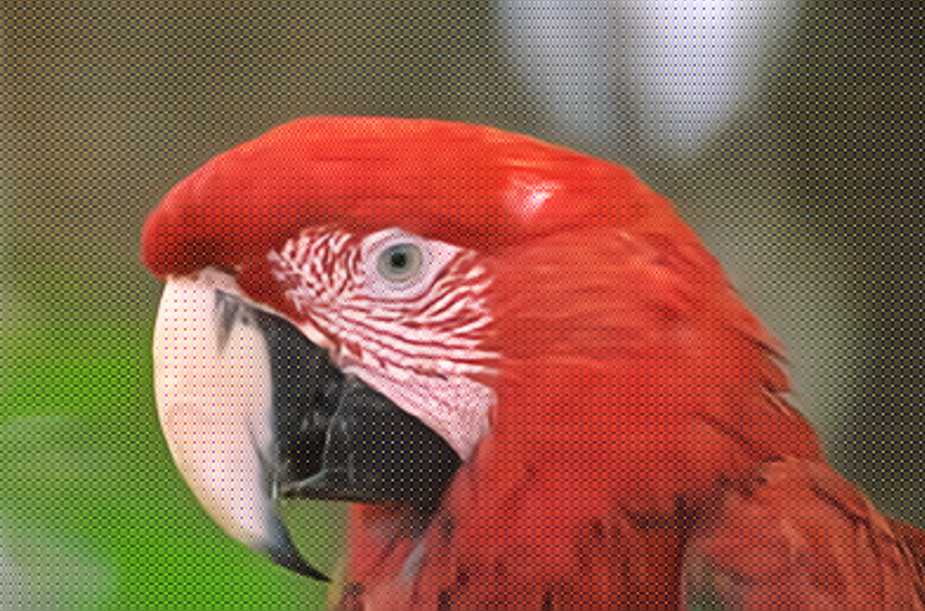}
        \caption{\centering \texttt{Mbt2018} \\ PSNR = 19.73 dB}
        \label{fig:VMAF_YUV_fine_tuning_img_mbt2018}
    \end{subfigure}
    \begin{subfigure}[b]{0.235\textwidth}
        \includegraphics[width=\linewidth]{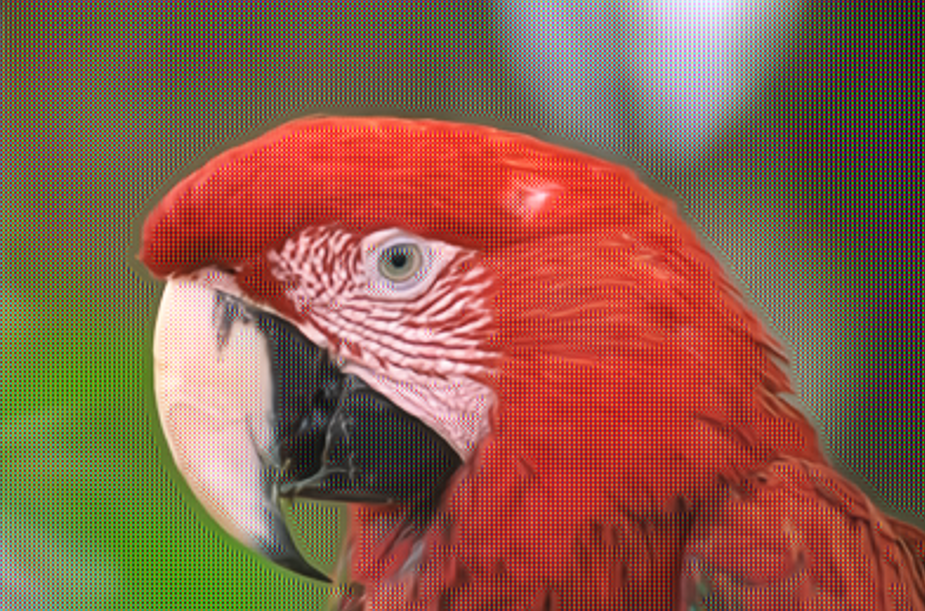}
        \caption{\centering \texttt{TCM} \\ PSNR = 12.23 dB}
        \label{fig:VMAF_YUV_fine_tuning_img_TCM}
    \end{subfigure}
    \caption{Samples from \textbf{fine-tuning decoders} using only VMAF NEG as distortion loss ($\alpha=\beta=0, \gamma=1$).}
    \label{fig:VMAF_YUV_fine_tuning}
\end{figure}

\begin{table}[t]
    \centering
    \begin{tabular}{c@{}c|r|r|r|r}
         & \multicolumn{1}{c|}{Block} & \multicolumn{1}{c|}{PSNR} & \multicolumn{1}{c|}{VMAF} & \multicolumn{1}{c|}{\#param} & \multicolumn{1}{c}{$\Delta$PSNR (dB) /} \\
         \multirow{3}{*}{\rotatebox{90}{\texttt{TCM} \hspace{2.5cm} \texttt{Mbt2018} \hspace{1.4cm} \texttt{Cheng2020} \hspace{1.1cm}}} & \multicolumn{1}{c|}{no.} & \multicolumn{1}{c|}{(dB)} & \multicolumn{1}{c|}{NEG Y} & \multicolumn{1}{c|}{[M]} & \multicolumn{1}{c}{1M params}\\
         \hline
         & none & 31.55 & 79.78 &  &  \\
         & 0 & 30.93 & 80.19 & 0.30 &  -2.13\\
         & 1 & 30.82 & 81.23 & 1.34 &  -0.55\\
         & 2 & 31.15 & 81.11 & 0.30 &  -1.35\\
         & 3 & 30.60 & 81.66 & 1.34 &  -0.71\\
         & 4 & 30.71 & 81.29 & 0.30 &  -2.87\\
         & 5 & 23.63 & 82.34 & 1.34 &  -5.90\\
         & 6 & 21.93 & 82.36 & 0.30 &  -32.61\\
         & 7 & 23.49 & 81.85 & 0.01 & -582.69 \\
         \hline
         & none & 31.68 & 78.75 & & \\
         & 0 & 30.94 & 80.50 & 0.92 & -0.81 \\
         & 1 & 30.92 & 80.35 & 0.04 & -20.37 \\
         & 2 & 30.67 & 80.85 & 0.92 & -1.09 \\
         & 3 & 30.96 & 80.71 & 0.04 & -19.30 \\
         & 4 & 24.34 & 81.55 & 0.92 & -7.96 \\
         & 5 & 30.86 & 80.69 & 0.04 & -22.20 \\
         & 6 & 27.29 & 80.58 & 0.01 & -304.97 \\
         \hline
         & none & 35.20 & 87.23 & & \\
         & 0 & 34.43 & 88.49 & 3.11 & -0.25 \\
         & 1 & 34.41 & 88.54 & 0.16 & -5.00 \\
         & 2 & 34.40 & 88.54 & 0.16 & -5.07 \\
         & 3 & 34.30 & 88.64 & 1.34 & -0.67 \\
         & 4 & 34.21 & 88.64 & 0.16 & -6.24 \\
         & 5 & 34.14 & 88.66 & 0.16 & -6.73 \\
         & 6 & 29.40 & 89.06 & 1.34 & -4.31 \\
         & 7 & 24.04 & 89.19 & 0.16 & -70.34 \\
         & 8 & 22.88 & 89.27 & 0.16 & -77.67 \\
         & 9 & 25.80 & 88.95 & 0.01 & -679.50 \\
    \end{tabular}
    \caption{Performance after \textbf{fine-tuning single decoder blocks} with only the VMAF NEG loss ($\alpha=\beta=0, \gamma=1$). The $\Delta \text{PSNR}$ is measured in dB relative to the baseline (none), and divided by the number of parameters in M. Fine-tuning was done for a single epoch. }
    \label{tab:block_fine_tuning}
\end{table}

\begin{table*}[h!]
    \centering
    \begin{tabular}{c@{}c|c|c|c|c|c|c|c|c|c|c|c|c|c}
        \multirow{3}{*}{\rotatebox{90}{Decoder \hspace{1.8cm} Encoder \hspace{2.4cm}}}
        \multirow{3}{*}{\rotatebox{90}{fine-tuning \hspace{1.5cm} fine-tuning \hspace{2.2cm}}} \\
        
         &&\multicolumn{3}{c|}{Loss factors} & \multicolumn{3}{|c|}{\texttt{Cheng2020}} & \multicolumn{3}{|c|}{\texttt{Mbt2018}} & \multicolumn{3}{|c|}{\texttt{TCM}} \\
         
         &&$\alpha$ & $\beta$ & $\gamma$ & PSNR & MS- & VMAF & PSNR & MS- & VMAF & PSNR & MS- & VMAF & Accum. \\
         &&&&& (dB) & SSIM & NEG Y &  (dB) & SSIM & NEG Y & (dB) & SSIM & NEG Y & rank \\
         \hline
         &&1.0 &  &  & 
         \textbf{\textcolor{green}{31.59}} & \textcolor{red}{0.971} & \textcolor{red}{79.70} & 
         \textbf{\textcolor{green}{31.68}} & \textcolor{red}{0.970} & \textcolor{red}{78.76} &  
         \textbf{\textcolor{green}{35.21}} & \textcolor{red}{0.985} & \textcolor{red}{87.01} & 
         39 \\ %MSE

         && & 1.0 &  & 
         \textcolor{red}{29.71} & \textbf{\textcolor{green}{0.977}} & \textbf{\textcolor{red}{76.27}} & 
         \textbf{\textcolor{red}{29.61}} & \textbf{\textcolor{green}{0.977}} & \textbf{\textcolor{red}{75.58}} & 
         \textcolor{red}{32.12} & \textbf{\textcolor{green}{0.987}} & \textbf{\textcolor{red}{84.31}} &
         \textbf{\textcolor{red}{49}} \\ %MS-SSIM
         
          &&& & 1.0 & 
         \textbf{\textcolor{red}{29.22}} & \textbf{\textcolor{red}{0.964}} & \textbf{\textcolor{green}{82.83}} &
         \textcolor{red}{29.94} & \textbf{\textcolor{red}{0.967}} & \textcolor{green}{82.48} & 
         \textbf{\textcolor{red}{31.91}} & \textbf{\textcolor{red}{0.980}} & \textbf{\textcolor{green}{89.37}} & 
         \textcolor{red}{45} \\ %VMAF NEG-YUV

         %  &&& 0.1 & 0.9 & 
         % 29.78 & 0.972 & 82.69 & 
         % 30.09 & 0.972 & 82.22 & 
         % 32.54 & 0.986 & 89.27 & 
         % 0 \\ %MIX MS-SSIM

         % &&0.1 &  & 0.9 & 
         % 30.88 & 0.968 & 82.92 &
         % 31.00 & 0.968 & 82.37 &
         % 34.36 & 0.984 & 89.43 & 
         % 0 \\ %MIX MSE

         &&0.8 & 0.1 & 0.1 & 
         \textcolor{green}{31.55} & 0.972 & 80.62 & 
         \textcolor{green}{31.63} & 0.972 & 79.96 & 
         \textcolor{green}{35.17} & 0.986 & 87.93 & 
         36 \\ %MIX3

         &&0.6 & 0.2 & 0.2 & 
         31.45 & 0.973 & 81.22 & 
         31.57 & 0.973 & 80.62 & 
         35.12 & 0.987 & 88.33 &
         33 \\ %MIX4

         &&0.4 & 0.3 & 0.3 & 
         31.31 & 0.974 & 81.57 & 
         31.42 & 0.974 & 80.95 & 
         34.97 & 0.987 & 88.66 &
         \textcolor{green}{31} \\ %MIX6

         &&0.2 & 0.4 & 0.4 & 
         31.10 & \textcolor{green}{0.975} & 81.65 & 
         31.26 & \textcolor{green}{0.975} & 81.26 &
         34.72 & 0.987 & 88.72 &
         \textbf{\textcolor{green}{29}} \\ %MIX7

         &&0.1 & 0.1 & 0.8 & 
         30.93 & 0.973 & \textcolor{green}{82.67} & 
         31.10 & 0.972 & \textbf{\textcolor{green}{83.03}} &
         34.42 & \textcolor{green}{0.986} & \textcolor{green}{89.24} & 
         34 \\ %MIX 2.2

         \hline \hline
         &&1.0 &  &  & 
         \textbf{\textcolor{green}{31.60}} & \textcolor{green}{0.970} & \textbf{\textcolor{red}{79.61}} & 
         \textbf{\textcolor{green}{31.71}} & \textcolor{green}{0.970} & \textbf{\textcolor{red}{78.72}} & 
         \textbf{\textcolor{green}{35.28}} & \textbf{\textcolor{green}{0.985}} & \textcolor{red}{87.42} &
         31 \\ %MSE

         && & 1.0 &  & 
         31.45 & \textbf{\textcolor{green}{0.971}} & \textcolor{red}{79.95} & 
         31.53 & \textbf{\textcolor{green}{0.971}} & \textcolor{red}{79.47} &
         35.13 & \textbf{\textcolor{green}{0.985}} & \textbf{\textcolor{red}{87.28}} & 
         \textbf{\textcolor{red}{40}} \\ %MS-SSIM
         
          &&&  & 1.0 & 
         \textbf{\textcolor{red}{13.86}} & \textbf{\textcolor{red}{0.538}} & \textbf{\textcolor{green}{85.10}} & 
         \textbf{\textcolor{red}{20.91}} & \textbf{\textcolor{red}{0.862}} & \textbf{\textcolor{green}{82.03}} & 
         \textbf{\textcolor{red}{13.61}} & \textbf{\textcolor{red}{0.931}} & \textbf{\textcolor{green}{89.88}} &
         \textcolor{red}{38} \\ %VMAF NEG-YUV

         &&0.8 & 0.1 & 0.1 & 
         \textcolor{green}{31.57} & \textcolor{green}{0.970} & \textcolor{green}{82.22} & 
         \textcolor{green}{31.69} & \textcolor{green}{0.970} & 79.60 & 
         \textcolor{green}{35.27} & \textbf{\textcolor{green}{0.985}} & 88.02 &
         \textbf{\textcolor{green}{25}} \\ %MIX3

         &&0.6 & 0.2 & 0.2 & 
         31.52 & \textcolor{green}{0.970} & 80.73 & 
         31.66 & \textcolor{green}{0.970} & 80.00 & 
         35.17 & \textbf{\textcolor{green}{0.985}} & 88.24 &
         \textcolor{green}{30} \\ %MIX4

         &&0.4 & 0.3 & 0.3 & 
         31.49 & \textcolor{green}{0.970} & 81.07 & 
         31.61 & \textcolor{green}{0.970} & 80.30 & 
         35.15 & \textbf{\textcolor{green}{0.985}} & 88.44 &
         \textcolor{green}{30} \\ %MIX6

         &&0.2 & 0.4 & 0.4 & 
         31.37 & \textcolor{green}{0.970} & 81.39 & 
         \textcolor{red}{31.52} & \textcolor{green}{0.970} & 80.61 &
         34.99 & \textbf{\textcolor{green}{0.985}} & 88.71 &
         32 \\ %MIX7

         &&0.1 & 0.1 & 0.8 & 
         \textcolor{red}{31.09} & \textcolor{red}{0.969} & 81.82 & 
         31.23 & \textcolor{red}{0.969} & \textcolor{green}{81.15} & 
         \textcolor{red}{34.65} & \textcolor{green}{0.984} & \textcolor{green}{88.95} &
         36 \\ %MIX 2.2

         \hline \hline 
         \multicolumn{5}{c|}{Best combined} &
         30.99 & 0.974 & 81.37 & 
         31.15 & 0.974 & 80.59 &
         34.58 & 0.987 & 88.75 &

    \end{tabular}
    \caption{Performance after \textbf{encoder fine-tuning} and \textbf{decoder fine-tuning} of the given codecs. Hyper-parameters $\alpha, \beta$, and $\gamma$ represent the share of MSE, MS-SSIM, and VMAF NEG in the distortion term of our proposed loss (\ref{ex:loss_function}).  We scored per metric and codec every mixed loss experiment. In the upper and middle table segments, loss factor combinations are separately ranked (1, 2, ..., 8). In "Accum. rank", we provide the ranks accumulated per row (lower is better). Best metric score is \textbf{\textcolor{green}{bold and green}}, second best is \textcolor{green}{green}, worst is \textbf{\textcolor{red}{bold and red}}, second worst is \textcolor{red}{red}. In the lower segment, both separately fine-tuned best ranked encoder and decoder are concatenated.}
    \label{tab:finetuning_final}
\end{table*}

For Table \ref{tab:first_fine_tuning_decoder}, we fine-tuned either the encoder (ENC) or the decoder (DEC) of our selected codecs with either MSE ([$\alpha, \beta, \gamma$] = [$1, 0, 0$]) or with the \texttt{VMAF-torch} version of VMAF NEG ([$\alpha, \beta, \gamma$] = [$0, 0, 1$]). 
We observe that fine-tuning the \textit{encoder} with VMAF NEG always results in a lower PSNR (red color) than the baseline and encoder fine-tuning with MSE. VMAF NEG scores, however, are better throughout (green color). When fine-tuning the \textit{decoder}, we see the worst PSNR scores across the table, along with better VMAF NEG scores than baseline and fine-tuning the decoder with MSE. The poor PSNR scores along with the strong VMAF NEG scores indicate a fine-tuning attack on VMAF NEG, which will be confirmed by subjective results in the following.

Figure \ref{fig:VMAF_YUV_fine_tuning} shows samples of our decoders being fine-tuned with VMAF NEG. In Fig.\ \ref{fig:VMAF_YUV_fine_tuning_img_cheng2020}, we clearly see two types of artifacts when fine-tuning \texttt{Cheng2020}: horizontal lines covering the image and ringing. Figs.\ \ref{fig:VMAF_YUV_fine_tuning_img_mbt2018} and \ref{fig:VMAF_YUV_fine_tuning_img_TCM}, which are from fine-tuned \texttt{Mbt2018} and \texttt{TCM} decoders, respectively, also show noticeable high-frequency overlaid checkerboard artifacts, when inspecting closer the image on a screen.

% \begin{figure*}[h]
%     \centering
%     \includegraphics[width=0.8\linewidth]{images/big_picture.pdf}
%     \caption{Samples from the \textbf{fine-tuned encoders} (top) and \textbf{fine-tuned decoders} (bottom).}
%     \label{fig:big_picture}
% \end{figure*}

\begin{figure*}[h]
    \centering
    \includegraphics[width=1\linewidth]{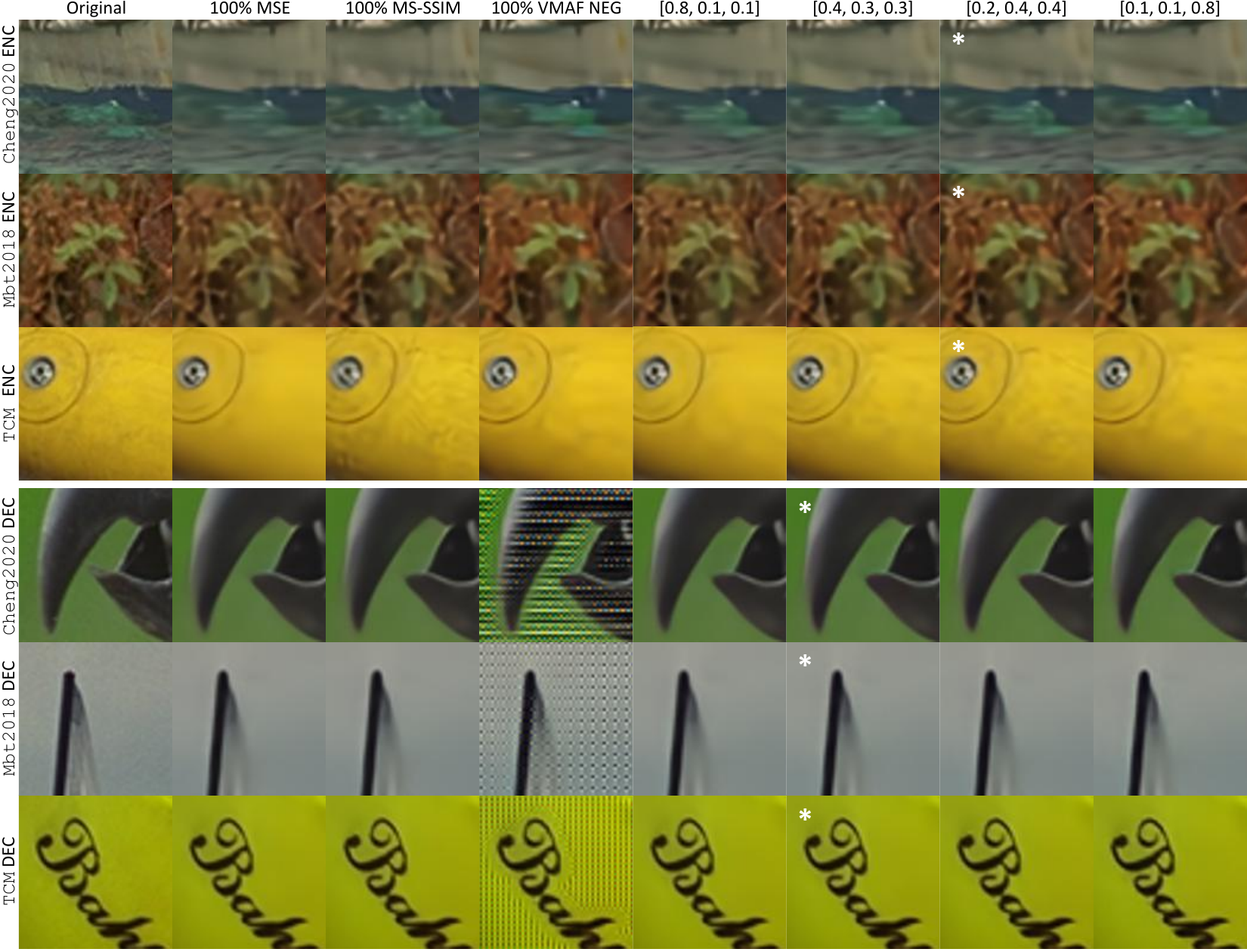}
    \caption{Patches from the \textbf{fine-tuned encoders} (top) and \textbf{fine-tuned decoders} (bottom). For better quality refers to the digital version. The white asterisk ($\ast$) marks the beset hyperparameter setting for encoder fine-tuning (top) or decoder fine-tuning (bottom), respectively. For better quality inspection, the reader is referred to the digital version on screen.}
    \label{fig:bigger_picture}
\end{figure*}

Table \ref{tab:block_fine_tuning} shows results from fine-tuning of only a \textit{single} block of the decoder with VMAF NEG ([$\alpha, \beta, \gamma$]=[$0, 0, 1$]), where 0 marks the earliest block and 7, 6, or 9 marking the last block of the decoder, respectively. From Table \ref{fig:VMAF_YUV_fine_tuning}, we already know that the PSNR drastically decreases by pure VMAF NEG fine-tuning.
Here, we measured the PSNR and VMAF NEG on the Y channel and provide the number of parameters in that block in millions. We also report on how much the PSNR (in dB) decrased \textit{per 1 M parameters} when fine-tuning that particular block. This gives an idea of how much effect the parameters in each block have to the distortion of the decoded images. 

We observe that fine-tuning \texttt{Cheng2020}'s layers 7 and 5 have similar impact on PSNR, but layer 7 has less then 1\% of the parameters of layer 5, therefore making it $>100$ times more sensitive. This can be seen by the drastically higher $\Delta$PSNR(dB) / 1 M params score of layer 7. In general, for all three codecs, there is a trend that later blocks have a stronger impact on PSNR than earlier blocks. These findings indicate that when decoding an image, the artifacts mainly stem from the last few blocks of the decoder.

Table \ref{tab:finetuning_final} shows the results of multiple fine-tunings of the \textit{encoders} and \textit{decoders}, according to our proposed loss (\ref{ex:loss_function}). The first row in the upper and the middle table segment is the baseline, individually for each codec, additionally MSE fine-tuned ($\alpha=1$) for a fair comparison. The best results per metric and codec are marked in green and bold, second in green, the worst in red and bold, the second worst in red. We ranked each column separately in each table segment from 1 to 8 and accumulated these ranks along each row to provide a representative accumulated rank for each loss configuration [$\alpha, \beta, \gamma$] in (\ref{ex:loss_function}), shown in the last column. A lower accumulated rank is better\footnote{\ninept Note that a similar averaging of single-metrics results has been successfully applied in the URGENT Challenge \cite{Challenge}.}. We pursued bitrate matching as described in Section \ref{sec:bitrate_matching}. The lower segment/row in Table \ref{tab:finetuning_final} shows the best fine-tuned encoder and the best fine-tuned decoder concatenated.

Taking first a bird's eye view onto Table \ref{tab:finetuning_final}, we realize that when optimizing either the encoder or the decoder for a single out of the three metrics (MSE, MS-SSIM, VMAF NEG), we obtain best or second best results in that particular metric, which is expected. A bit surprising, however, is the poor performance on the other two metrics in this case. For decoder fine-tuning, the situation is similar but optimizing for MS-SSIM ($\beta = 1$) at least produces respectable PSNR results. However, when optimizing for VMAF NEG only ($\gamma = 1$), both for encoder and decoder fine-tuning, we see our earlier results confirmed that the metrics PSNR and now also MS-SSIM show the worst results among all evaluated methods, confirming VMAF NEG's fine-tuning attack. Apart from the VMAF NEG-only optimization, we observe very similar (good) results in the MS-SSIM metric both for encoder and decoder fine-tuning.

Now let's analyze the fine-tunings with our proposed loss (\ref{ex:loss_function}) a bit deeper. Among the encoder fine-tunings, [$\alpha, \beta, \gamma$] = [$0.2, 0.4, 0.4$] is the clear winner with an accumulated rank of only 29, compared to the best single-metrics loss (MSE) with 39. It shows a balanced performance without any poor results among each of the three investigated codecs. Among the decoder fine-tunings, [$\alpha, \beta, \gamma$] = [$0.8, 0.1, 0.1$] is the clear winner with an accumulated rank of only 25, compared to the best single-metric loss (MSE) with 31. Different to MSE-only optimization ($\alpha=1.0$, three times worst or second worst performance), the decoder fine-tuning winner is 7 out of 9 times the best or second best, without any red-marked low metrics result, accordingly, also showing a much more balanced quality. We observe that for encoder fine-tuning, the MSE loss portion plays a minor role ($\alpha = 0.2$), whereas its optimal weight in the decoder fine-tuning ($\alpha = 0.8$) is much higher. However, we see the significant and positive effect of mixing MS-SSIM and VMAF NEG into (\ref{ex:loss_function}) by choosing $\beta=\gamma=0.1$ in this case.

In the lowest segment (row) of Table \ref{tab:finetuning_final} we investigate a mixed-world scenario, where the content is re-encoded with the improved fine-tuned encoder [0.2, 0.4, 0.4], and decoded by the already improved fine-tuned decoders [0.8, 0.1, 0.1]. Note that this is not a joint fine-tuning. We observe that most of the time the combination of best fine-tuned encoder and best fine-tuned decoder has no benefit over the strong encoder fine-tuning (ENC). In comparison to the best fine-tuned decoder (DEC), operating on a standard encoder, the combined fine-tunings are ahead in MS-SSIM and mostly in VMAF NEG Y as well.

In Figure \ref{fig:bigger_picture}, we show cropped samples from the Kodak dataset for qualitative visual verification. The top half depicts decoded images from codecs with fine-tuned encoders, while the bottom half those with fine-tuned decoders. We display samples from most of the different loss approaches seen in Table \ref{tab:finetuning_final} for each of the three codecs. \footnote{\ninept The images from encoder fine-tuning (ENC) lie within the following ranges: \texttt{Cheng2020}: 0.33 bpp $\pm 1.7\%$, \texttt{Mbt2018}: 0.49 bpp $\pm 2.1\%$, \texttt{TCM}: 0.81 bpp $\pm 1.6\%$. For the decoder fine-tuning (DEC), due to the fixed encoder, the bitrates are exactly the same.}

Starting with the \texttt{Cheng2020} encoder, two prominent features that change a lot depending on the loss function are the yellow stain in the top right corner and the light reflection on the water in the middle of the image. For the \texttt{Mbt2018} encoder, it is mainly the saturation of the leafs in the middle of the image and the background. The image patches from the \texttt{TCM} encoder mainly change in their visibility of the stains on the rubber boat and the overall color of the rubber boat. From all these changes the ones on the rubber boat seem to be the most noticeable ones.

When looking at those features and especially at the well noticeable distortions from the \texttt{TCM} encoder experiment, we find that our fine-tuning with [$\alpha, \beta, \gamma$] = [$0.2, 0.4, 0.4$] brought the best results regarding visual quality and fidelity in comparison to the ground truth. This is in line with our quantitative findings from Table \ref{tab:finetuning_final} where this method also received the best overall score.
 
As in Figure \ref{fig:VMAF_YUV_fine_tuning}, we see in Figure \ref{fig:bigger_picture} that for fine-tuning the decoder (DEC) of any of our given codecs with the VMAF NEG loss only, the artifacts reappear. Besides declaring the 100\% VMAF NEG approach to be the worst method to fine-tune the decoder with, we don't find one method to be superior to any of the others. We also see in Table \ref{tab:finetuning_final} that except for fine-tuning with [$\alpha, \beta, \gamma$] = [$0.2, 0.4, 0.4$], the deviation of our metrics when fine-tuning the decoder are much smaller then when fine-tuning the encoder.

We find this confirmed by the following investigation: Excluding the 100\% VMAF NEG fine-tuning due to its obvious flaws, the \textit{mean} (over all codecs) $\text{metrics \textit{range}} =\max - \min$  (per codec and metric) when fine-tuning the encoder is 2.35 dB for PSNR, while it is only 0.44 dB for the decoder fine-tuning. For MS-SSIM, we find 0.005 (ENC) and 0.002 (DEC). When looking at VMAF NEG, it is 6.93 (ENC) and only 2.24 (DEC). These results confirm that the variance when fine-tuning the decoder (excluding 100\% VMAF NEG fine-tuning) is much smaller than when fine-tuning the encoder. Accordingly, fine-tuning the decoder turns out to be less sensitive as long as any of the $\alpha, \beta, \gamma$ is not too dominant. This much smaller variance of the decoder fine-tuning also explains the not so noticeable DEC differences in Figure \ref{fig:bigger_picture} when excluding the 100\% VMAF NEG approach.

\vfill\pagebreak
\section{Conclusions}
\label{sec:Conclusions}
In our work we have shown and analyzed pitfalls during optimization of various image encoders or decoders towards the VMAF NEG metric. Without employing a proxy neural network or any tedious adversarial training, we proposed a fine-tuning loss consisting of three differentiable metrics, i.e., MSE, MS-SSIM, but also VMAF NEG. It turned out that an appropriate mix of several metrics delivers overall best ranks, and, equally important, very balanced results: For encoder fine-tuning, the optimal share of VMAF NEG and MS-SSIM in the loss function is 80\% (only 20\% for MSE), whereas for decoder fine-tuning, it is just vice versa: MSE takes a share of 80\% and the remaining 20\% are for VMAF NEG and MS-SSIM. Having performed investigations with three different learned image codecs, our findings show good generalization over the investigated various codecs.

\vfill\pagebreak
%\ninept
\sloppy
\bibliographystyle{IEEEtran}
\bibliography{refs} % Greift auf deine refs.bib zu (ohne die Endung .bib geschrieben!)

\end{document}